\documentclass[aps,prl,twocolumn,showpacs]{revtex4}
\usepackage{graphicx,amsmath,amssymb}
\begin{document}

\renewcommand{\b}[1]{\mathbf{#1}}
\renewcommand{\c}[1]{\mathcal{#1}}
\renewcommand{\u}{\uparrow}
\renewcommand{\d}{\downarrow}
\newcommand{\bsigma}{\boldsymbol{\sigma}}
\newcommand{\blambda}{\boldsymbol{\lambda}}
\newcommand{\tr}{\mathop{\mathrm{tr}}}
\newcommand{\arcsinh}{\mathop{\mathrm{arcsinh}}}
\newcommand{\sgn}{\mathop{\mathrm{sgn}}}
\newcommand{\half}{{\textstyle\frac{1}{2}}}
\newcommand{\sh}{{\textstyle{\frac{1}{2}}}}
\newcommand{\ish}{{\textstyle{\frac{i}{2}}}}
\newcommand{\thf}{{\textstyle{\frac{3}{2}}}}
\newcommand{\um}{\mu\text{m}}

\title{Kondo effect in the helical edge liquid of the quantum spin Hall state}

\author{Joseph Maciejko$^{1}$, Chaoxing Liu$^{1,2}$, Yuval Oreg$^{3}$,
Xiao-Liang Qi$^1$, Congjun Wu$^4$, and Shou-Cheng Zhang$^1$}

\affiliation{$^1$ Department of Physics, McCullough Building,
Stanford University, Stanford, CA 94305-4045, USA}

\affiliation{$^2$ Center for Advanced Study, Tsinghua University,
Beijing, 100084, China}

\affiliation{$^3$ Department of Condensed Matter Physics, Weizmann
Institute of Science, Rehovot 76100, Israel}

\affiliation{$^4$ Department of Physics, University of California,
San Diego, CA 92093, USA}

\date\today

\begin{abstract}
Following the recent observation of the quantum spin Hall (QSH)
effect in HgTe quantum wells, an important issue is to understand
the effect of impurities on transport in the QSH regime. Using
linear response and renormalization group methods, we calculate
the edge conductance of a QSH insulator as a function of
temperature in the presence of a magnetic impurity. At high
temperatures, Kondo and/or two-particle scattering give rise to a
logarithmic temperature dependence. At low temperatures, for weak
Coulomb interactions in the edge liquid the conductance is
restored to unitarity with unusual power-laws characteristic of a
`local helical liquid', while for strong interactions transport
proceeds by weak tunneling through the impurity where only half an
electron charge is transferred in each tunneling event.
\end{abstract}

\pacs{
71.55.-i,    % impurity and defect levels
72.25.Hg,           % electrical injection of spin polarized carriers
73.43.-f,           % quantum Hall effects
75.30.Hx            % magnetic impurity interactions
}

\maketitle

The quantum spin Hall (QSH) insulator is a topologically
non-trivial state of matter \cite{Konig2008} that has recently
been observed in transport experiments carried out in HgTe quantum
wells (QW) \cite{Konig2007} following its theoretical prediction
\cite{Bernevig2006}. The two-dimensional QSH insulator has a
charge excitation gap in the bulk but supports one-dimensional
gapless edge states forming a so-called helical liquid (HL): on
each edge there exists a Kramers' pair of counter-propagating
states with opposite spin polarization. The QSH insulator is
robust against weak single-particle perturbations which preserve
time-reversal symmetry such as weak potential scattering
\cite{Kane2005,Wu2006,Xu2006}.

This theoretical picture is consistent with experimental
observations: the longitudinal conductance $G$ in a Hall bar
measurement is approximately quantized to $G_0=2e^2/h$,
independent of temperature, for samples of about a micron length
\cite{Konig2007,MarkusThesis}. However, larger samples exhibit
$G<G_0$ and $G$ decreases with decreasing temperature
\cite{MarkusThesis}. Deviations from the expected quantized value
have been attributed to the presence of local doped regions due to
potential inhomogeneities within the sample arising from
impurities or roughness of the well/barrier interface
\cite{MarkusThesis}. Although pure potential scattering cannot
backscatter the edge states, the role of these potential
inhomogeneities is to trap bulk electrons which may then interact
with the edge electrons. These localized regions act as dephasing
centers for the edge channels due to interaction effects and may
cause backscattering.

\begin{figure}[h]
\begin{center}
\includegraphics[width=2in,angle=-90]{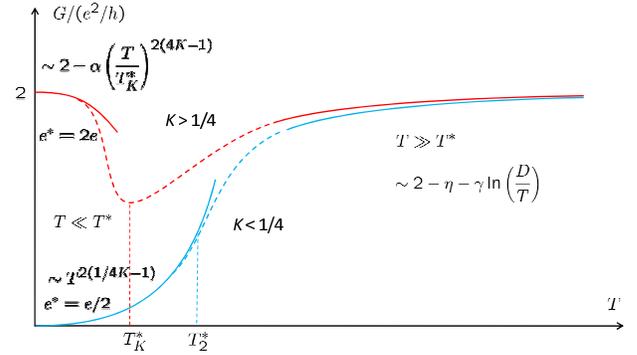}
\end{center}
\caption{Temperature dependence of the conductance: the behavior
is
 logarithmic at high temperature $T\gg T^*$ and power-law at low
  temperature $T\ll T^*$. At $T=0$, a metal-insulator quantum phase transition is driven by Coulomb interactions in the helical
   liquid: the system is a `Kondo metal' for $K>1/4$ and a
    `Luttinger liquid insulator' for $K<1/4$. The Fano factor $e^*$ is defined
    as the ratio between shot noise and current, and reflects the
    charge of the current-carrying excitations.}\label{fig:strong}
\end{figure}

In this work, we study theoretically the temperature dependence of
the edge conductance of a QSH insulator. We consider the case
where a local doped region in the vicinity of the edge contains an
odd number of electrons and acts as a magnetic impurity coupled to
the helical edge liquid. Our main results are as follows (Fig.
\ref{fig:strong}):

1. At high temperatures, the conductance is
    logarithmic, $-\Delta G\equiv-(G-G_0)=\eta+\gamma\ln(D/T)$ where $\eta,\gamma$ are
interaction-dependent parameters and $D$ is an energy scale of
order the bulk gap.

2. For weak Coulomb interactions
    $K>1/4$ where $K$ is the Luttinger parameter of the edge liquid,
    the conductance is restored to unitarity at $T=0$ due to the formation of a Kondo singlet. This is in stark
    contrast with the Kondo problem in a usual spinful 1D
    liquid where the conductance vanishes
    at $T=0$ for all $K_\rho<1$ where $K_\rho$ is the Luttinger
    parameter in the charge sector \cite{furusaki1994}.
    At low but finite $T$ the conductance decreases as
    an
    unusual power-law $\Delta G\propto-T^{2(4K-1)}$ due to
    correlated
    two-particle (2P)
     backscattering. The edge liquid being helical, the decrease
      in conductance is a direct
     measure of the spin-flip rate \cite{Garst2005}.

3. For strong Coulomb interactions
    $K<1/4$, 2P backscattering processes are relevant and the
    system becomes insulating at $T=0$. At low but finite $T$, the
    conductance is restored by tunneling of excitations with
    fractional charge $e/2$ and we obtain
    $G(T)\propto T^{2(1/4K-1)}$.

\emph{Model.}---We model the impurity by a $S=\frac{1}{2}$ local
spin coupled by exchange interaction to the 1D HL with Coulomb
interactions. The HL having the same number of degrees of freedom
as a spinless fermion, a single nonchiral boson $\phi$ is
sufficient for its description in the bosonized language
\cite{Wu2006}. The system is described by the Hamiltonian
$H=H_0+H_\mathrm{K}+H_2$ where $H_0$ is the usual
Tomonaga-Luttinger Hamiltonian $H_0=\frac{v}{2}\int
dx\left[K\Pi^2+\frac{1}{K}(\partial_x\phi)^2\right]$, with $K$ the
Luttinger parameter and $v$ the edge state velocity. The Kondo
Hamiltonian $H_\mathrm{K}$ has the form
\begin{eqnarray}
H_\mathrm{K}=\frac{J_\parallel a}{2\pi\xi}\bigl(S_-\colon
e^{-i2\sqrt{\pi}\phi(0)}
\colon+\mathrm{h.c.}\bigr)-\frac{J_za}{\sqrt{\pi}}S_z\Pi(0),\label{Model_HK}
\end{eqnarray}
where $S_\pm=S_x\pm iS_y$ and $S_z$ are the spin operators for the
impurity localized at $x=0$. $a$ is the lattice constant of the
underlying 2D lattice and corresponds to the size of the impurity
(we assume that the impurity occupies a single lattice site).
$\xi$, the penetration length of the helical edge states into the
bulk, acts as a short-distance cutoff for the 1D continuum theory
in the same way that the magnetic length, the penetration length
of the chiral edge states, acts as a short-distance cutoff for the
chiral Luttinger liquid theory of the quantum Hall edge
\cite{Wen1991}. In addition to Kondo scattering, 2P backscattering
is allowed by time-reversal symmetry \cite{Wu2006,Xu2006}. In HgTe
QW the wavevector at which the edge dispersion enters the bulk is
usually much smaller than $\pi/2a$ such that the uniform 2P
backscattering (umklapp) term requiring $4k_F=2\pi/a$ can be
ignored. The impurity potential can however provide a $4k_F$
momentum transfer and we must generally also consider a local
impurity-induced 2P backscattering term \cite{Meidan2005}
$H_2=\frac{\lambda_2a^2}{2\pi^2\xi^2}\colon\cos4\sqrt{\pi}\phi(0)\colon$
where $\lambda_2$ is the 2P backscattering amplitude.

\emph{Weak coupling regime.}---We first consider the weak coupling
regime where $J_\parallel$, $J_z$ and $\lambda_2$ are small
parameters. The calculation of the conductance proceeds in two
steps. We first perform an explicit perturbative calculation of
the conductance to quadratic order in the bare couplings
$J_\parallel$ and $\lambda_2$ using the Kubo formula. This result
is then extended to include all leading logarithmic terms in the
perturbation expansion by means of a weak coupling renormalization
group (RG) analysis of the scale-dependent couplings
$J_\parallel(T)$ and $\lambda_2(T)$ where the scale is set by the
temperature $T$.

The forward scattering term $J_z$ can be removed from the
Hamiltonian by a unitary transformation \cite{Emery1992} of the
form $U=e^{i\lambda S_z\phi(0)}$ with
$\lambda=-J_za/vK\sqrt{\pi}$, at the expense of modifying the
scaling dimension of the vertex operator $\colon
e^{i2\sqrt{\pi}\phi}\colon$. The transformed Hamiltonian reads
$UHU^\dag=H_0+H_2+\frac{J_\parallel a}{2\pi \xi} \left(S_-\colon
e^{-i2\sqrt{\pi}\chi\phi(0)}\colon+\mathrm{h.c.}\right)$, where
$\chi=1-\nu J_z/2K$ with $\nu=a/\pi v$ the density of states of
the HL. The scaling dimension of $\colon
e^{i2\sqrt{\pi}\chi\phi}\colon$ is $\tilde{K}\equiv K\chi^2$.

Using the transformed Hamiltonian we obtain the correction to the
conductance $\Delta G(T)=\Delta G_\mathrm{K}(T)+\Delta G_2(T)$ to
quadratic order in the couplings $J_\parallel$ and $\lambda_2$
using the Kubo formula, where $\Delta G_\mathrm{K}$ is the
correction due to spin-flip Kondo scattering \cite{footnote1},
\begin{equation}
\frac{\Delta G_\mathrm{K}}{e^2/h}=-\frac{\Gamma(\half)
\Gamma(\tilde{K})}{\Gamma(\half+\tilde{K})}\frac{\pi^2}{3}S(S+1)
(\nu J_\parallel(T))^2,\label{WC_GK}
\end{equation}
where $\nu J_\parallel(T)=\nu J_\parallel(T/D)^{\tilde{K}-1}$ to
$\c{O}(J_\parallel)$ and $D=\hbar v/\pi\xi$ is a high-energy
cutoff of order the bulk gap. $\Delta G_2$ is the correction due
to 2P backscattering,
\begin{eqnarray}
    \frac{\Delta G_2}{e^2/h}=-\frac{\Gamma(\half)\Gamma(4K)}
    {\Gamma(\half+4K)}\frac{a^4}{2\pi^4\xi^4}
    \left(\frac{\lambda_2(T)}{D}
    \right)^2,
    \label{WC_G2}
\end{eqnarray}
where $\lambda_2(T)=\lambda_2(T/D)^{4K-1}$. There are no crossed
terms of the form $\c{O}(J_\parallel\lambda_2)$ or
$\c{O}(J_z\lambda_2)$ for a $S=\frac{1}{2}$ impurity since the 2P
backscattering operator flips two spins but a $S=\frac{1}{2}$ spin
can be flipped only once. We however expect that such terms would
be generated for impurities with higher spin.

These results can be complemented by a RG analysis. The RG
equation for $\lambda_2$ follows by dimensional analysis
$\frac{d\lambda_2}{d\ell}=(1-4K)\lambda_2$ with $\ell=\ln(D/T)$,
so that $\lambda_2$ is relevant for $K<1/4$ and irrelevant for
$K>1/4$. The renormalized coupling is
$\lambda_2(T)=\lambda_2(T/D)^{4K-1}$, and second order
renormalized perturbation theory $\Delta G_2(T)\propto
-\lambda_2(T)^2$ simply reproduces the Kubo formula result
(\ref{WC_G2}). Perturbation theory fails for $T\lesssim T_2^*$
where $T_2^*\propto(\lambda_2^0)^{1/(1-4K)}$ is a scale for the
crossover from weak to strong 2P backscattering with $\lambda_2^0$
the bare 2P backscattering amplitude.

The one-loop RG equations \cite{Schiller1995,Wu2006} for the Kondo
couplings $J_\parallel,J_z$ read
\begin{eqnarray}
    \frac{dJ_\parallel}{d\ell}=(1-K)J_\parallel+\nu J_\parallel J_z,\hspace{5mm}\frac{dJ_z}{d\ell}=\nu J_\parallel^2.
    \label{WC_RG_J}
\end{eqnarray}
The family of RG trajectories is indexed by a single scaling
invariant $c=(\nu J_\parallel)^2-(\nu \tilde{J}_z)^2$ where
$\nu\tilde{J}_z\equiv\nu J_z+1-K$, which is fixed by the couplings
at energy scale~$D$. In contrast to the spinful case
\cite{furusaki1994}, the absence of spin-flip forward scattering
in the HL preserves the stability of the ferromagnetic fixed line,
as is the case in the usual Kondo problem. The renormalized
spin-flip amplitude $J_\parallel(T)$ is given in terms of the bare
parameters $J_\parallel^0,J_z^0$ by
\begin{equation}\label{nuJ}
\nu J_\parallel(T)=\frac{\alpha\nu J_\parallel^0}
{\sinh\left(\alpha\nu J_\parallel^0\ln(T/T_\mathrm{K}^*)\right)},
\end{equation}
such that $\Delta G_\mathrm{K}$ is obtained to all orders in
perturbation theory in the leading-log approximation by
substituting Eq. (\ref{nuJ}) in Eq. (\ref{WC_GK}). In Eq.
(\ref{nuJ}),
$\alpha=\bigl[(\tilde{J}_z^0/J_\parallel^0)^2-1\bigr]^{1/2}$ is an
anisotropy parameter \cite{footnote2} and $T_\mathrm{K}^*$ is the
Kondo temperature,
\[
T_\mathrm{K}^*=D\exp\left(-\frac{1}{\nu J_\parallel^0}\frac{\arcsinh\alpha}
{\alpha}\right).
\]
In the isotropic case $\alpha=0$, one recovers the usual form
$T_\mathrm{K}^*= De^{-1/\nu J_\parallel^0}$. In the limit $1-K\gg
\nu J_\parallel^0,\nu J_z^0$, Coulomb interactions dominate over
Kondo physics and we obtain $T_\mathrm{K}^*\simeq D\left(\frac{\nu
J_\parallel^0}{1-K} \right)^{1/(1-K)}$, a power-law dependence
similar to results previously obtained \cite{Lee1992} for Kondo
impurities in spinful Luttinger liquids. From the scaling exponent
we see that $T_\mathrm{K}^*$ corresponds to the scale of the mass
gap opened in a spinless Luttinger liquid by a point potential
scatterer of strength $\nu J_\parallel^0$ and the corresponding
crossover is that of weak to strong single-particle
backscattering.

In the high temperature limit $\max\{T_2^*,T_\mathrm{K}^*\}\ll
T\lesssim D$, both the Kondo and 2P scattering processes
contribute logarithmically to the suppression of the conductance,
$-\Delta G(T)\sim\eta+\gamma\ln(D/T)$ where $\eta,\gamma$ are
functions of the bare couplings $K$, $J^0_\parallel,J^0_z$ and
$\lambda_2^0$.
\begin{figure}[t]
%\begin{center}
\includegraphics[angle=180,width=3.5in]{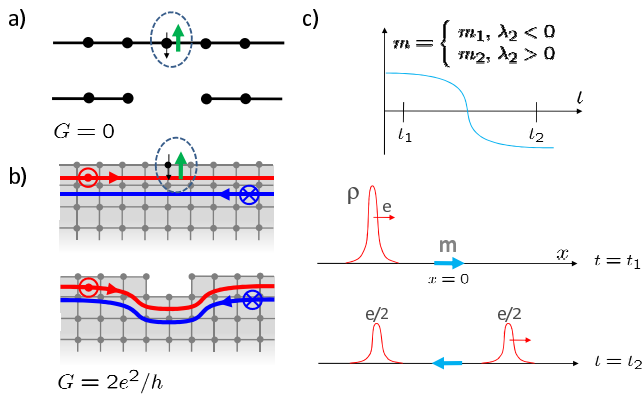}
\vspace{10mm}
%\end{center}
\caption{Strong coupling regime: the Kondo singlet effectively
removes one site from the system. a) Luttinger liquid with $K_\rho<1$: a punctured
1D lattice is disconnected, b) QSH edge liquid with $K>1/4$: the edge liquid
follows the boundary of the deformed 2D lattice. c) Half-charge tunneling for $K<1/4$ by flips of the Ising order parameter $m$.}\label{fig:edge}
\end{figure}

\emph{Strong coupling regime.}---We now investigate the low
temperature regime below the crossover temperatures
$T\ll\min\{T_2^*,T_\mathrm{K}^*\}$. The topological nature of the
QSH edge state as a `holographic liquid' living on the boundary of
a 2D system \cite{Wu2006} results in a drastic change of the
low-energy effective theory in the vicinity of the strong coupling
fixed point as compared to that of a usual 1D quantum wire. As
suggested by the perturbative RG analysis, the nature of the $T=0$
fixed point depends on whether $K$ is greater or lesser than
$1/4$.

We first consider the case where $K>1/4$. In this case, 2P
backscattering is irrelevant and $\Delta G_2$ flows to zero. On
the other hand, for antiferromagnetic $J_z$ the Kondo strong
coupling fixed point $J_\parallel,J_z\rightarrow+\infty$ is
reached at $T=0$, with formation of a local Kramers singlet and
complete screening of the impurity spin by the edge electrons. As
a result, the formation of the Kondo singlet effectively removes
the impurity site from the underlying 2D lattice (Fig.
\ref{fig:edge}b). In a strictly 1D spinful liquid, this has the
effect of cutting the system into two disconnected semi-infinite
1D liquids (Fig. \ref{fig:edge}a) and transport is blocked at
$T=0$ for all $K_\rho<1$ \cite{furusaki1994}. In contrast, due to
its topological nature, the QSH edge state simply follows the new
shape of the edge and we expect the unitarity limit $G=2e^2/h$ to
be restored at $T=0$. For finite $T\ll T_\mathrm{K}^*$, the
effective low-energy Hamiltonian contains the leading irrelevant
operators in the vicinity of the fixed point. In the case of
spinful conduction electrons, the lowest-dimensional operator
causing a reduction of the conductance is single-particle
backscattering. However, the helical property of the QSH edge
states forbids such a term and it is natural to conjecture that
the leading irrelevant operator must be the 2P backscattering
operator with scaling dimension $4K$. We thus expect a correction
to the conductance at low temperatures $T\ll T_\mathrm{K}^*$ for
$K>1/4$ of the form $\Delta
G\propto-\left(T/T_\mathrm{K}^*\right)^{2(4K-1)}$.

In particular, in the noninteracting case $K=1$ we predict a $T^6$
dependence in marked contrast to both the usual Fermi liquid
\cite{Nozieres1974} and spinful 1D liquid \cite{furusaki1994}
behaviors. This dependence characteristic of a `local helical
Fermi liquid' can be understood from a simple phase space
argument. The Pauli principle requires the 2P backscattering
operator to be defined through a point-splitting procedure
\cite{Wu2006} with the short-distance cutoff $\xi$, which
translates into a derivative coupling in the limit of small $\xi$,
\[
\psi_R^\dag(0)\psi_R^\dag(\xi)\psi_L(0)\psi_L(\xi)
\rightarrow
\xi^2\psi_R^\dag\partial_x\psi_R^\dag\psi_L\partial_x\psi_L.
\]
In the absence of derivatives, the four fermion term contributes
$T^2$ to the inverse lifetime $\tau_k^{-1}$. The derivatives
correspond to four powers of momenta close to the Fermi points in
the scattering rate $\Gamma_{k,k'\rightarrow
p,p'}\propto(k-k')^2(p-p')^2$, which translates into an additional
factor of $T^4$. Furthermore, since at temperatures $T\ll
T^*_\mathrm{K}$ suppression of the conductance is entirely due to
two-electron scattering, we expect the effective charge $e^*\equiv
S/(2|\langle I_B\rangle|)$ obtained from a measurement of the shot
noise $S$ in the backscattering current $\langle I_B\rangle$ to be
$e^*=2e$ \cite{Meidan2005,footnote4}.

For $K<1/4$, the $|\lambda_2|\rightarrow\infty$ fixed point is
reached at $T=0$ and the system becomes insulating. The field
$\phi(x=0,\tau)$ is pinned at the minima of the cosine potential
$H_2$ located at $\pm(2n+1)\sqrt{\pi}/4$ for $\lambda_2>0$ and
$\pm 2n\sqrt{\pi}/4$ for $\lambda_2<0$, with $n\in\mathbb{Z}$. The
conductance is restored at finite temperatures by instanton
processes corresponding to the tunneling between nearby minima
separated by $\Delta\phi=\sqrt{\pi}/2$. From the relation
$j_e=e\partial_t\phi/\sqrt{\pi}$ between electric current $j_e$
and $\phi$ field, the charge pumped by a single instanton (Fig.
\ref{fig:edge}c) is obtained as $\Delta
Q_\mathrm{inst}=\frac{e}{\sqrt{\pi}}\Delta \phi=\frac{e}{2}$. This
fractionalized tunneling current can be understood as the
Goldstone-Wilczek current \cite{goldstone1981,Qi2008b,Qi2008} for
1D Dirac fermions with a mass term
$\delta\c{L}=g\bar{\Psi}(m_1+i\gamma^5m_2)\Psi$ where
$g\sim\lambda_2$, and the mass order parameters
$m_1=\cos2\sqrt{\pi}\phi$, $m_2=\sin2\sqrt{\pi}\phi$ change sign
during an instanton process with $\Delta\phi=\sqrt{\pi}/2$. The
order is Ising-like because the 2P backscattering term explicitly
breaks the spin $U(1)$ symmetry of the helical liquid
$H_0=\frac{\pi v}{2}\int dx(K\sigma_z^2+\frac{1}{K}\rho^2)$ down
to $\mathbb{Z}_2$, where $\sigma_z=\rho_+-\rho_-$ and
$\rho=\rho_++\rho_-$ are the spin and charge densities, and
$\rho_\pm$ are the chiral densities for the two members of the
Kramers pair.

Fractionalization of the tunneling current is confirmed by a
saddle-point evaluation of the path integral for large $\lambda_2$
in the dilute instanton gas approximation, which yields a Coulomb
gas representation of the partition function that can be mapped
exactly to the boundary sine-Gordon theory
\[
S[\theta]=\frac{K}{\beta}\sum_{i\omega_n}|\omega_n||\theta(i\omega_n)|^2+t\int_0^\beta
d\tau\,\cos\sqrt{\pi}\theta(\tau),
\]
where $t$ is the instanton fugacity. The RG equation for $t$
follows as $\frac{dt}{d\ell}=\left(1-\frac{1}{4K}\right)t$ and the
conductance $G(T)\propto t(T)^2$ is a power-law
$G\propto\left(T/T_2^*\right)^{2(1/4K-1)}$ for $T\ll
T_2^*,\,K<1/4$. In contrast to the strong coupling regime in a
usual Luttinger liquid where $t$ corresponds to a single-particle
hopping amplitude \cite{Kane1992}, the unusual scaling dimension
of the tunneling operator in the present case corresponds to
half-charge tunneling. In particular, we calculate the shot noise
in the strong coupling regime using the Keldysh approach
\cite{Martin2005} and find $S=2e^*|\langle I\rangle|$ where
$\langle I\rangle$ is the tunneling current and $e^*=e/2$.

\emph{Experimental realization.}---We find that the experimental
results of Ref. \cite{MarkusThesis} are consistent with our
theoretical expressions for the weak coupling regime with a weak
Luttinger parameter $K\simeq 1$, but the small number of available
data points does not allow for a reliable determination of the
model parameters. The temperature dependence of the conductance
being exponentially sensitive to $K$, our predictions can be best
verified in QW with stronger interaction effects. Due to reduced
screening of the Coulomb interaction \cite{footnote3}, we expect
to see a steeper decrease of conductance with decreasing
temperature in HgTe samples with only a backgate.

Because of lower Fermi velocities $v_F$, we expect even stronger
interaction effects to occur in InAs/GaSb/AlSb type-II QW
\cite{Cooper1998} which have been recently predicted to exhibit
the QSH effect \cite{Liu2008}. For QW widths
$w_\mathrm{InAs}=w_\mathrm{GaSb}=10$ nm in the inverted regime
\cite{Liu2008}, and considering only screening from the front gate
closest to the QW layer, from a $\b{k}\cdot\b{p}$ calculation of
material parameters we obtain $K\simeq 0.2<1/4$, making the
insulating phase observable at low temperatures. Although the
backgate will cause additional screening, $v_F$ can be further
decreased by adding a thin AlSb barrier layer between the InAs and
GaSb QW layers. The Fermi velocity is controlled by the overlap
between electron and hole subband wavefunctions
\cite{Bernevig2006} which are localized in different QW layers in
the type-II configuration \cite{Liu2008}, and an additional
barrier layer will decrease this overlap. A lower $v_F$ also
translates into higher Kondo temperatures since $\nu J\propto
1/v_F^2$, where one power of $v_F$ comes from the matrix element
of the localized impurity potential between edge states, and one
power comes from the density of states $\nu$. Since
$T_\mathrm{K}^*$ depends on $\nu J$ exponentially, we expect
experimentally accessible Kondo temperatures in type-II QW.

We are grateful to E.-A. Kim, T. Hughes, M. K\"{o}nig, E. Berg, T.
Ong, A. Furusaki, H. Yao and S. Raghu for valuable discussions.
J.M. is supported by the National Science and Engineering Research
Council (NSERC) of Canada, the Fonds qu\'{e}b\'{e}cois de la
recherche sur la nature et les technologies (FQRNT), and the
Stanford Graduate Fellowship Program. C.X.L. is supported by the
CSC, NSF under grant numbers 10774086, 10574076, and Basic
Research Development of China under grant number 2006CB921500.
Y.O. is supported by BSF, DIP and GIF grants, the Mel Schwartz
Research Fund, an NSF CAREER award DMR-0349354, and the Stanford
Institute for Theoretical Physics (SITP). C.W. is supported by the
NSF under grant number DMR-0804775, the Sloan Research Foundation,
and the Army Research Office under grant number W911NF0810291.
S.C.Z. is supported by the U.S. Department of Energy, Office of
Basic Energy Sciences under contract DE-AC03-76SF00515.

\bibliography{impurity0330}

\begin{thebibliography}{27}
\expandafter\ifx\csname natexlab\endcsname\relax\def\natexlab#1{#1}\fi
\expandafter\ifx\csname bibnamefont\endcsname\relax
  \def\bibnamefont#1{#1}\fi
\expandafter\ifx\csname bibfnamefont\endcsname\relax
  \def\bibfnamefont#1{#1}\fi
\expandafter\ifx\csname citenamefont\endcsname\relax
  \def\citenamefont#1{#1}\fi
\expandafter\ifx\csname url\endcsname\relax
  \def\url#1{\texttt{#1}}\fi
\expandafter\ifx\csname urlprefix\endcsname\relax\def\urlprefix{URL }\fi
\providecommand{\bibinfo}[2]{#2}
\providecommand{\eprint}[2][]{\url{#2}}

\bibitem[{\citenamefont{\textrm{For a review, see M. K\"{o}nig} \emph{et
  al.}}(2008)}]{Konig2008}
\bibinfo{author}{\bibnamefont{\textrm{For a review, see M. K\"{o}nig} \emph{et
  al.}}}, \bibinfo{journal}{J. Phys. Soc. Jpn} \textbf{\bibinfo{volume}{77}},
  \bibinfo{pages}{031007} (\bibinfo{year}{2008}).

\bibitem[{\citenamefont{\textrm{M. K\"{o}nig} \emph{et al.}}(2007)}]{Konig2007}
\bibinfo{author}{\bibnamefont{\textrm{M. K\"{o}nig} \emph{et al.}}},
  \bibinfo{journal}{Science} \textbf{\bibinfo{volume}{318}},
  \bibinfo{pages}{766} (\bibinfo{year}{2007}).

\bibitem[{\citenamefont{Bernevig et~al.}(2006)\citenamefont{Bernevig, Hughes,
  and Zhang}}]{Bernevig2006}
\bibinfo{author}{\bibfnamefont{B.~A.} \bibnamefont{Bernevig}},
  \bibinfo{author}{\bibfnamefont{T.~L.} \bibnamefont{Hughes}},
  \bibnamefont{and} \bibinfo{author}{\bibfnamefont{S.~C.} \bibnamefont{Zhang}},
  \bibinfo{journal}{Science} \textbf{\bibinfo{volume}{314}},
  \bibinfo{pages}{1757} (\bibinfo{year}{2006}).

\bibitem[{\citenamefont{Kane and Mele}(2005)}]{Kane2005}
\bibinfo{author}{\bibfnamefont{C.~L.} \bibnamefont{Kane}} \bibnamefont{and}
  \bibinfo{author}{\bibfnamefont{E.~J.} \bibnamefont{Mele}},
  \bibinfo{journal}{Phys. Rev. Lett.} \textbf{\bibinfo{volume}{95}},
  \bibinfo{pages}{226801} (\bibinfo{year}{2005}).

\bibitem[{\citenamefont{Wu et~al.}(2006)\citenamefont{Wu, Bernevig, and
  Zhang}}]{Wu2006}
\bibinfo{author}{\bibfnamefont{C.}~\bibnamefont{Wu}},
  \bibinfo{author}{\bibfnamefont{B.~A.} \bibnamefont{Bernevig}},
  \bibnamefont{and} \bibinfo{author}{\bibfnamefont{S.~C.} \bibnamefont{Zhang}},
  \bibinfo{journal}{Phys. Rev. Lett.} \textbf{\bibinfo{volume}{96}},
  \bibinfo{pages}{106401} (\bibinfo{year}{2006}).

\bibitem[{\citenamefont{Xu and Moore}(2006)}]{Xu2006}
\bibinfo{author}{\bibfnamefont{C.}~\bibnamefont{Xu}} \bibnamefont{and}
  \bibinfo{author}{\bibfnamefont{J.~E.} \bibnamefont{Moore}},
  \bibinfo{journal}{Phys. Rev. B} \textbf{\bibinfo{volume}{73}},
  \bibinfo{pages}{045322} (\bibinfo{year}{2006}).

\bibitem[{\citenamefont{\textrm{M. K\"{o}nig}}(2007)}]{MarkusThesis}
\bibinfo{author}{\bibnamefont{\textrm{M. K\"{o}nig}}}, Ph.D. thesis,
  \bibinfo{school}{\textrm{University of W\"{u}rzburg}} (\bibinfo{year}{2007}).

\bibitem[{\citenamefont{Furusaki and Nagaosa}(1994)}]{furusaki1994}
\bibinfo{author}{\bibfnamefont{A.}~\bibnamefont{Furusaki}} \bibnamefont{and}
  \bibinfo{author}{\bibfnamefont{N.}~\bibnamefont{Nagaosa}},
  \bibinfo{journal}{Phys. Rev. Lett.} \textbf{\bibinfo{volume}{72}},
  \bibinfo{pages}{892} (\bibinfo{year}{1994}).

\bibitem[{\citenamefont{Garst et~al.}(2005)\citenamefont{Garst, \textrm{P.
  W\"{o}lfle}, Borda, von Delft, and Glazman}}]{Garst2005}
\bibinfo{author}{\bibfnamefont{M.}~\bibnamefont{Garst}},
  \bibinfo{author}{\bibnamefont{\textrm{P. W\"{o}lfle}}},
  \bibinfo{author}{\bibfnamefont{L.}~\bibnamefont{Borda}},
  \bibinfo{author}{\bibfnamefont{J.}~\bibnamefont{von Delft}},
  \bibnamefont{and} \bibinfo{author}{\bibfnamefont{L.}~\bibnamefont{Glazman}},
  \bibinfo{journal}{Phys. Rev. B} \textbf{\bibinfo{volume}{72}},
  \bibinfo{pages}{205125} (\bibinfo{year}{2005}).

\bibitem[{\citenamefont{Wen}(1991)}]{Wen1991}
\bibinfo{author}{\bibfnamefont{X.-G.} \bibnamefont{Wen}},
  \bibinfo{journal}{Phys. Rev. B} \textbf{\bibinfo{volume}{44}},
  \bibinfo{pages}{5708} (\bibinfo{year}{1991}).

\bibitem[{\citenamefont{Meidan and Oreg}(2005)}]{Meidan2005}
\bibinfo{author}{\bibfnamefont{D.}~\bibnamefont{Meidan}} \bibnamefont{and}
  \bibinfo{author}{\bibfnamefont{Y.}~\bibnamefont{Oreg}},
  \bibinfo{journal}{Phys. Rev. B} \textbf{\bibinfo{volume}{72}},
  \bibinfo{pages}{121312(R)} (\bibinfo{year}{2005}).

\bibitem[{\citenamefont{Emery and Kivelson}(1992)}]{Emery1992}
\bibinfo{author}{\bibfnamefont{V.~J.} \bibnamefont{Emery}} \bibnamefont{and}
  \bibinfo{author}{\bibfnamefont{S.}~\bibnamefont{Kivelson}},
  \bibinfo{journal}{Phys. Rev. B} \textbf{\bibinfo{volume}{46}},
  \bibinfo{pages}{10812} (\bibinfo{year}{1992}).

\bibitem[{foo({\natexlab{a}})}]{footnote1}
\bibinfo{note}{This result is valid in the high-temperature regime $\hbar
  v/L\ll T<D$ for Fermi liquid leads where $L$ is the length of the QSH region
  (see Ref. \onlinecite{Maslov1995b}).}

\bibitem[{\citenamefont{Schiller and Ingersent}(1995)}]{Schiller1995}
\bibinfo{author}{\bibfnamefont{A.}~\bibnamefont{Schiller}} \bibnamefont{and}
  \bibinfo{author}{\bibfnamefont{K.}~\bibnamefont{Ingersent}},
  \bibinfo{journal}{Phys. Rev. B} \textbf{\bibinfo{volume}{51}},
  \bibinfo{pages}{4676} (\bibinfo{year}{1995}).

\bibitem[{foo({\natexlab{b}})}]{footnote2}
\bibinfo{note}{One can show that the Kondo model derived from the Anderson
  model for a single level coupled to the HL is isotropic due to time-reversal
  symmetry with
  $J_\parallel^0=J_z^0=(|t|^2+|u|^2)\left(\frac{1}{\epsilon_F-\epsilon_d}
  +\frac{1}{\epsilon_d+U-\epsilon_F}\right)$ where $\epsilon_F$ is the Fermi
  energy, $\epsilon_d$ is the impurity level with on-site Coulomb repulsion
  $U$, and $u$ and $t$ are the spin-flip and non-spin-flip hopping amplitudes,
  respectively. Coulomb interactions ($K\neq 1$) may however induce an
  effective anisotropy ($\alpha\neq 0$) even if the original Kondo model is
  isotropic.}

\bibitem[{\citenamefont{Lee and Toner}(1992)}]{Lee1992}
\bibinfo{author}{\bibfnamefont{D.-H.} \bibnamefont{Lee}} \bibnamefont{and}
  \bibinfo{author}{\bibfnamefont{J.}~\bibnamefont{Toner}},
  \bibinfo{journal}{Phys. Rev. Lett.} \textbf{\bibinfo{volume}{69}},
  \bibinfo{pages}{3378} (\bibinfo{year}{1992}).

\bibitem[{\citenamefont{Nozi\`{e}res}(1974)}]{Nozieres1974}
\bibinfo{author}{\bibfnamefont{P.}~\bibnamefont{Nozi\`{e}res}},
  \bibinfo{journal}{J. Low. Temp. Phys.} \textbf{\bibinfo{volume}{17}},
  \bibinfo{pages}{31} (\bibinfo{year}{1974}).

\bibitem[{foo({\natexlab{c}})}]{footnote4}
\bibinfo{note}{In the weak coupling or high temperature regime $T\gg
  T_2^*,T_\mathrm{K}^*$, both the Kondo ($e^*=e$) and 2P backscattering
  ($e^*=2e$) contributions to the effective carrier charge are present, such
  that we expect a non-universal value for the Fano factor.}

\bibitem[{\citenamefont{Goldstone and Wilczek}(1981)}]{goldstone1981}
\bibinfo{author}{\bibfnamefont{J.}~\bibnamefont{Goldstone}} \bibnamefont{and}
  \bibinfo{author}{\bibfnamefont{F.}~\bibnamefont{Wilczek}},
  \bibinfo{journal}{Phys. Rev. Lett.} \textbf{\bibinfo{volume}{47}},
  \bibinfo{pages}{986} (\bibinfo{year}{1981}).

\bibitem[{\citenamefont{Qi et~al.}(2008{\natexlab{a}})\citenamefont{Qi, Hughes,
  and Zhang}}]{Qi2008b}
\bibinfo{author}{\bibfnamefont{X.~L.} \bibnamefont{Qi}},
  \bibinfo{author}{\bibfnamefont{T.~L.} \bibnamefont{Hughes}},
  \bibnamefont{and} \bibinfo{author}{\bibfnamefont{S.~C.} \bibnamefont{Zhang}},
  \bibinfo{journal}{Nature Phys.} \textbf{\bibinfo{volume}{4}},
  \bibinfo{pages}{273} (\bibinfo{year}{2008}{\natexlab{a}}).

\bibitem[{\citenamefont{Qi et~al.}(2008{\natexlab{b}})\citenamefont{Qi, Hughes,
  and Zhang}}]{Qi2008}
\bibinfo{author}{\bibfnamefont{X.~L.} \bibnamefont{Qi}},
  \bibinfo{author}{\bibfnamefont{T.~L.} \bibnamefont{Hughes}},
  \bibnamefont{and} \bibinfo{author}{\bibfnamefont{S.~C.} \bibnamefont{Zhang}},
  \bibinfo{journal}{Phys. Rev. B} \textbf{\bibinfo{volume}{78}},
  \bibinfo{pages}{195424} (\bibinfo{year}{2008}{\natexlab{b}}).

\bibitem[{\citenamefont{Kane and Fisher}(1992)}]{Kane1992}
\bibinfo{author}{\bibfnamefont{C.~L.} \bibnamefont{Kane}} \bibnamefont{and}
  \bibinfo{author}{\bibfnamefont{M.~P.~A.} \bibnamefont{Fisher}},
  \bibinfo{journal}{Phys. Rev. B} \textbf{\bibinfo{volume}{46}},
  \bibinfo{pages}{15233} (\bibinfo{year}{1992}).

\bibitem[{\citenamefont{Martin}(2005)}]{Martin2005}
\bibinfo{author}{\bibfnamefont{T.}~\bibnamefont{Martin}},
  \bibinfo{journal}{ArXiv: cond-mat/0501208}  (\bibinfo{year}{2005}).

\bibitem[{foo({\natexlab{d}})}]{footnote3}
\bibinfo{note}{$K$ can be estimated \cite{Kane1992} by
  $K=[1+\alpha\ln(d/\ell)]^{-1/2}$ where
  $\alpha=\frac{2}{\pi^2}\frac{e^2/\epsilon}{\hbar v_F}$ and $\epsilon$ is the
  bulk dielectric constant. The distance $d$ from the QW layer to a nearby
  metallic gate acts as a screening length for the Coulomb potential, and
  $\ell$ is a microscopic length scale $\ell=\max\{\xi,w\}$ which acts as a
  short-distance cutoff for the Coulomb potential, where $w$ is the thickness
  of the QW layer.}

\bibitem[{\citenamefont{Cooper et~al.}(1998)\citenamefont{Cooper, Patel,
  Drouot, Linfield, Ritchie, and Pepper}}]{Cooper1998}
\bibinfo{author}{\bibfnamefont{L.~J.} \bibnamefont{Cooper}},
  \bibinfo{author}{\bibfnamefont{N.~K.} \bibnamefont{Patel}},
  \bibinfo{author}{\bibfnamefont{V.}~\bibnamefont{Drouot}},
  \bibinfo{author}{\bibfnamefont{E.~H.} \bibnamefont{Linfield}},
  \bibinfo{author}{\bibfnamefont{D.~A.} \bibnamefont{Ritchie}},
  \bibnamefont{and} \bibinfo{author}{\bibfnamefont{M.}~\bibnamefont{Pepper}},
  \bibinfo{journal}{Phys. Rev. B} \textbf{\bibinfo{volume}{57}},
  \bibinfo{pages}{11915} (\bibinfo{year}{1998}).

\bibitem[{\citenamefont{Liu et~al.}(2008)\citenamefont{Liu, Hughes, Qi, Wang,
  and Zhang}}]{Liu2008}
\bibinfo{author}{\bibfnamefont{C.}~\bibnamefont{Liu}},
  \bibinfo{author}{\bibfnamefont{T.~L.} \bibnamefont{Hughes}},
  \bibinfo{author}{\bibfnamefont{X.~L.} \bibnamefont{Qi}},
  \bibinfo{author}{\bibfnamefont{K.}~\bibnamefont{Wang}}, \bibnamefont{and}
  \bibinfo{author}{\bibfnamefont{S.~C.} \bibnamefont{Zhang}},
  \bibinfo{journal}{Phys. Rev. Lett.} \textbf{\bibinfo{volume}{100}},
  \bibinfo{pages}{236601} (\bibinfo{year}{2008}).

\bibitem[{\citenamefont{Maslov}(1995)}]{Maslov1995b}
\bibinfo{author}{\bibfnamefont{D.~L.} \bibnamefont{Maslov}},
  \bibinfo{journal}{Phys. Rev. B} \textbf{\bibinfo{volume}{52}},
  \bibinfo{pages}{{R}14368} (\bibinfo{year}{1995}).

\end{thebibliography}

\end{document}